\documentclass[review]{elsarticle}
\usepackage{lineno,hyperref}
\modulolinenumbers[5]

\usepackage{xcolor}
\usepackage{ragged2e}
\usepackage{graphicx}
\usepackage{pbox}
\usepackage{amsmath}
\usepackage{tabularx}
\usepackage{subfig}
\usepackage{graphicx}
\usepackage{amsmath, amsfonts, bm, color, ulem}
\usepackage{units}

\usepackage{comment}

\usepackage[english]{babel}

\makeatletter
\def\ps@pprintTitle{%
  \let\@oddhead\@empty
  \let\@evenhead\@empty
  \def\@oddfoot{\reset@font\hfil\thepage\hfil}
  \let\@evenfoot\@oddfoot
}
\makeatother

\begin{document}

\begin{frontmatter}

\title{Hilbert space representation of binary operations on a power-multiplying oscillator}

\author[1]{Elena Campillo Abarca}
\author[1]{Almudena Martínez Cedillo}
\author[1]{Jimena de Hita Fernández}
\author[1]{Miguel León Pérez}
\author[1]{Laura Morón Conde}
\author[1]{Andrei Sipos}
\author[1]{Daniel Heredia Doval}
\author[1]{Javier Domingo Serrano}
\author[1]{Rubén González Martínez}

\address[1]{BioCoRe, Biological Cooperative Research S.Coop, Madrid, Spain}



\begin{abstract}

In this study, the properties of an oscillating system composed of a pendulum connected to a seesaw and placed on a moving platform with a certain slope are analyzed. Using complex numbers to collect the information contained in the system proves to be crucial in order to observe the properties described by both cross and dot products. The representation of physical quantities in complex numbers reveals that for certain angles, precisely those where the oscillation translates into a displacement, the properties of the system are expressed in the real plane. 

\end{abstract}
\end{frontmatter}

\section{Introduction}

In this work the result of applying complex numbers to an oscillatory system with a certain slope and freedom of motion on a single coordinate axis will be presented. The use of complex numbers allows, by calculating the cross product and the dot product of certain vector magnitudes, to formalize mathematically certain behaviors of the system.

The earliest known written evidence of the calculation of a square root of a negative quantity dates from about 75 B.C. This description appeared in the book \textit{Stereometry}, written by the Greek Heron of Alexandria \cite{maluendas2breve}. It was in the $12^{th}$ century when, for the first time, the Hindu mathematician Bhaskara Acharya referred to the non-existence of the square root of a negative number \cite{agarwal2011history}. In the $16^{th}$ century, Girolamo Cardano collected in his book \textit{Ars Magna} the description of algebraic methods to solve cubic and quartic equations \cite{agarwal2011history,grossman2008algebra}.

The term imaginary for these quantities was coined by René Descartes in the Third Book his \textit{Geometry}, published in 1637 \cite{merino2006short}. However, it was not until the $18^{th}$ century when Leonhard Euler introduced a specific nomenclature for solving roots of negative numbers \cite{merino2006short}, and Friedrich Gauss completed the construction of a new numerical set, that of complex numbers \cite{maluendas2breve,merino2006short}.

\section{Methods}

The kinematic and dynamic behavior of an oscillator consisting of a pendulum connected to a seesaw and coupled to on a moving platform with a certain slope has been analyzed in previous works \cite{2021energy,2021twodimensional}.
In those previous studies, the forces of weight and inertia in the system were interpreted in the space $\mathbb{R}$. Now, in order to calculate the cross product and dot product the same forces will be given in the complex plane $\mathbb{C}$. The use of complex numbers to represent the system allows to elaborate a more accurate description of the system's behavior.
Therefore, we will begin by describing the system using the tool of complex numbers. Next, the cross product and dot product will be performed for certain properties of the system. And finally, the implications of the results obtained will be studied.

\section{Results}

%cálculos matemáticos del producto escalar y producto vectorial. Tanto la importancia numérica, como los gráficos, como las magnitudes

\subsection{Description of the system}

\begin{figure} [htb]
\centering
\includegraphics[width=80mm]{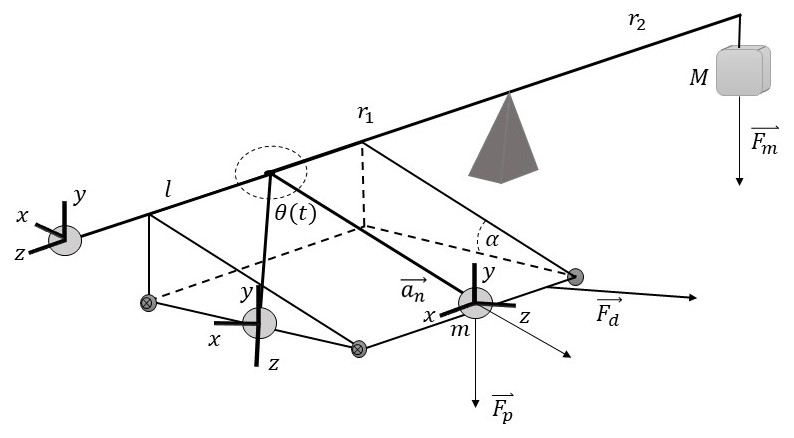}
\caption{Full oscillator description. System formed by a simple pendulum connected to a seesaw with arms $r_1$ and $r_2$. At the end of arm $r_1$ there is a pendulum of mass $m$ with a normal acceleration ($\vec{a}_n$) rotating about $\theta=360^\circ$ and with an inclination $\alpha$. $\vec{F_d}$ indicates the direction of displacement of the mobile. At the end of the arm $r_2$ there is the variable mass $M$ with weight $\vec{F_m}$ \cite{2021twodimensional}.}
\label{Fig1}
\end{figure}

To begin with, the description of the system will be expressed in the same terms as in previous works \cite{2021energy,2021twodimensional}. This system consists of a mobile with an inclined platform to which a pendulum composed of a mass $m$ and a fixed rod of length $l$ is attached. Thus, the pendulum swings at angle $\theta$ on a platform inclined at angle $\alpha$ (Fig.\ref{Fig1}). the platform counts with four wheels that enable it to move freely in a single coordinate in both its positive and negative axes. In this work, any frictional forces are neglected.

Additionally, the assumed system has, connected to the pendulum and in equilibrium relation, a seesaw of variable mass $M$, which adapts to the resultant of the set of forces operating at each angle $\theta$ of the oscillation. In previous studies \cite{2021energy,2021twodimensional} it has been observed that the torque at both ends is compensated and the relationship between the two masses and the angular configuration of the system has been formulated (Eq.\ref{eq:1}). Applying the formula to all possible values of $\theta$ and $\alpha$ the representative transformation matrix of the system was obtained (Eq.\ref{eq:2}). 

\begin{equation}
\frac{M}{m}=4\cos \theta \sin ^{2}\alpha +1
\label{eq:1}
\end{equation}

\begin{equation}
\frac{M}{m}\hspace{-0.05cm}=
\resizebox{0.83\hsize}{!}{%
$\displaystyle
\begin{pmatrix}
1-\frac{1}{2}(e^{i\theta}+e^{-i\theta})(e^{-i\alpha}-e^{i\alpha})^2 & 
\cdots &
1-\frac{1}{2}(e^{i(\theta+m)}+e^{-i(\theta+m)})(e^{-i\alpha}-e^{i\alpha})^2 \\
\vdots & 
\ddots & 
\vdots \\
1-\frac{1}{2}(e^{i\theta}+e^{-i\theta})(e^{-i(\alpha+n)}-e^{i(\alpha+n)})^2 &
\cdots  &
1-\frac{1}{2}(e^{i(\theta+m)}+e^{-i(\theta+m)})(e^{-i(\alpha+n)}-e^{i(\alpha+n)})^2
\end{pmatrix}
$}
\label{eq:2}
\end{equation}
\vspace{3mm}

As shown in Figure \ref{Fig1}, it is the wheels of the mobile itself that limit the displacement of the system to the $z$-axis. Therefore, the system behaves like a flywheel, releasing its energy only when the pendulum is in the same direction as the $z$-coordinate, i.e., when the kinetic energy of the pendulum's motion is at its maximum.

\subsection{Description of the system in complex numbers}

%habría que describirlo de otra forma, de tal manera que el sistema se vea descrito en espacio unitario

A vector space over the body $\mathbb{R}$ or $\mathbb{C}$ endowed with a scalar product is called a pre-Hilbert space. If it is also complete in the associated metric topology, it is said to be a Hilbert space \cite{abellanas1987espacios}. If the dimension is finite and the body is that of the real numbers, it will be said to be a Euclidean space; if the body is that of the complex numbers, and the dimension is finite, it will be said to be a unitary space \cite{munkres2018elements}. 

It is not possible to formalize mathematically the system's behavior in spherical coordinates because the motion is not bounded to a sphere of length given by the pendulum's rod, but is described for any spatial topology that is not reducible to the sphere determined by the rod. Thus, in order to formalize this topology it is proposed to represent the system in $\mathbb{C}$. The canonical basis in unit space can be written as \{(1,0,0),(0,1,0),(0,0,i)\}.

\begin{figure}[htb!]
\centering
\includegraphics[width=\linewidth]{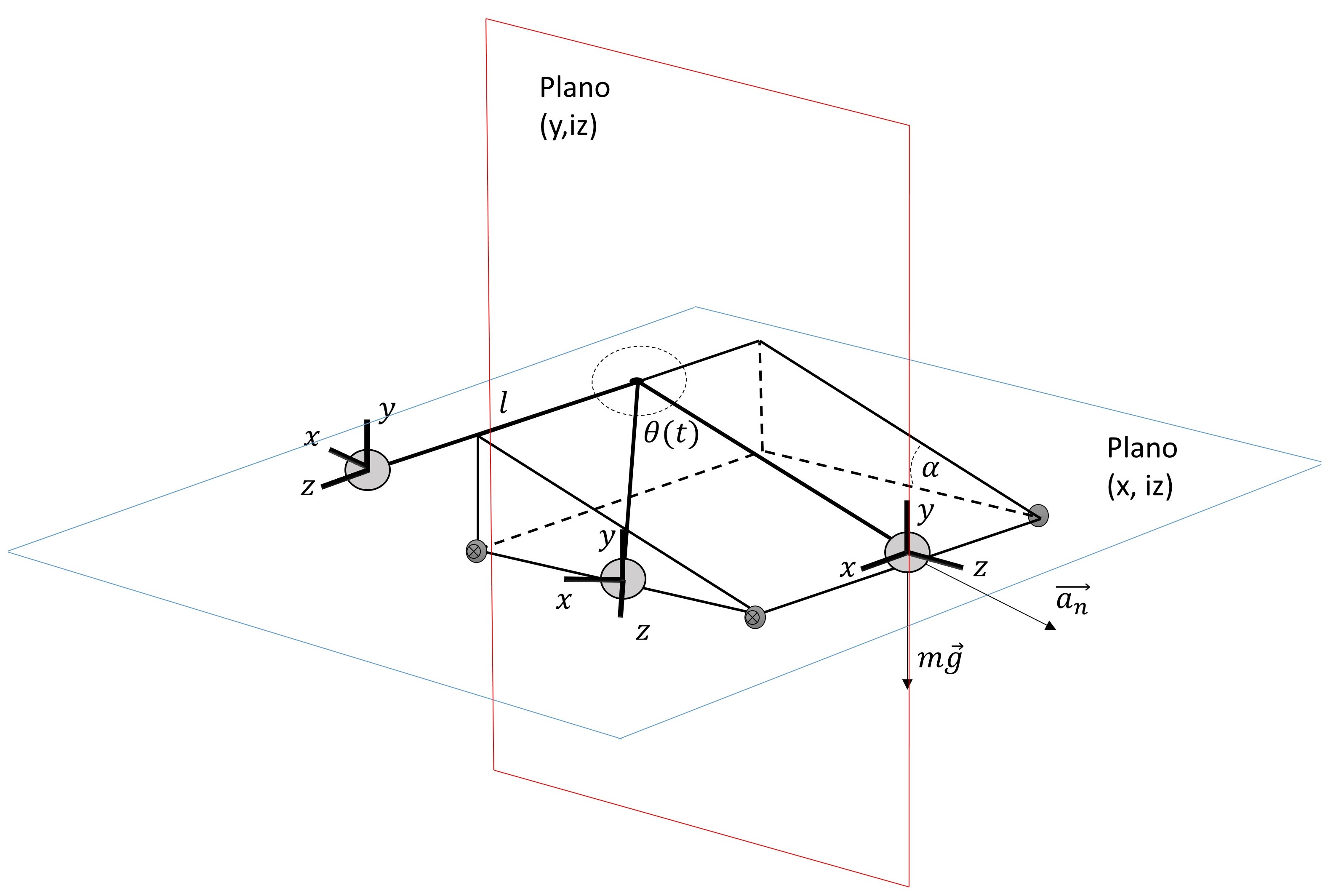}
\caption{Planes of the canonical basis. The real axes $x$ and $y$ and the imaginary axis $\mathrm{i}z$ are displayed.}
\label{Fig2}
\end{figure}

\begin{figure}[htb!]
\centering
\includegraphics[width=120mm]{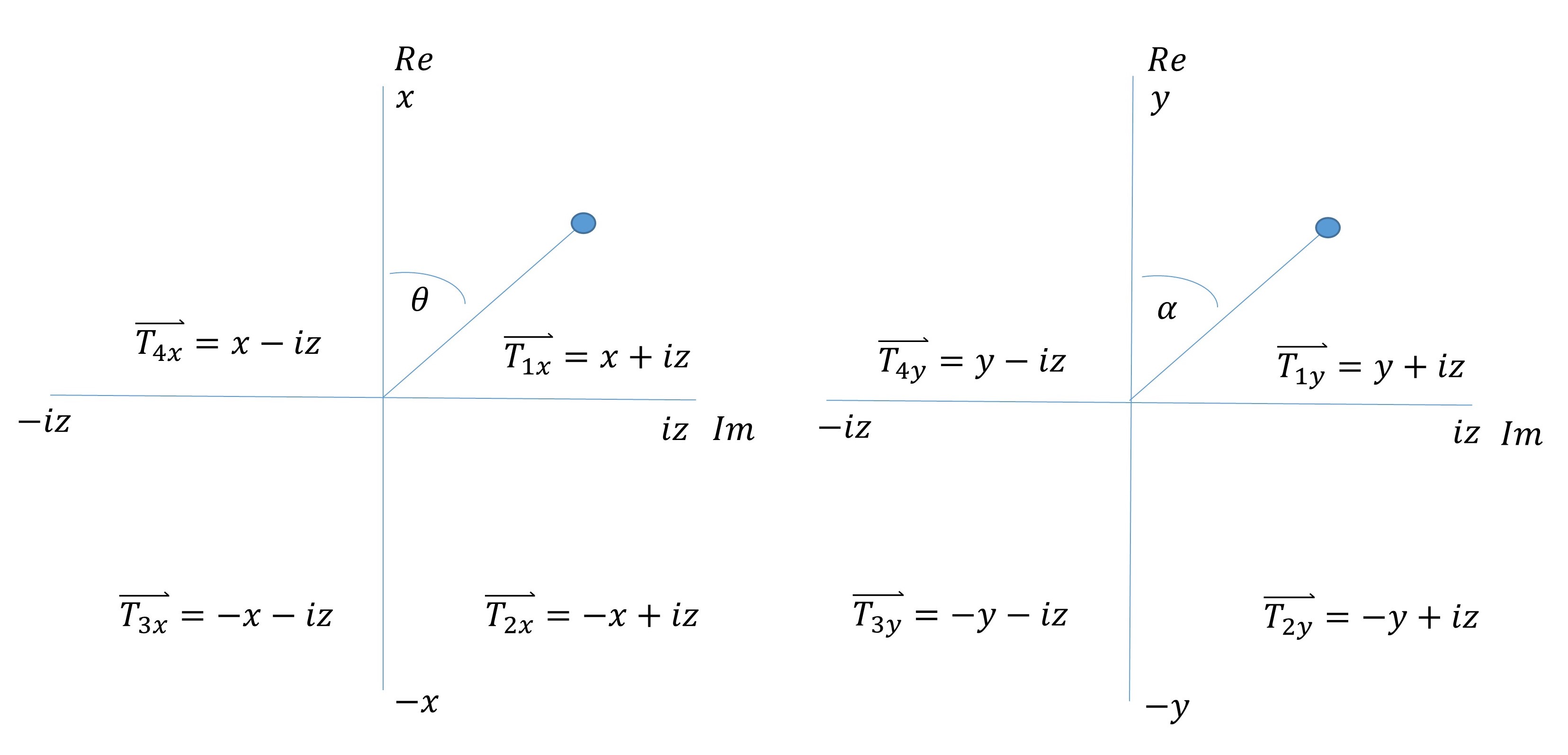}
\caption{Description of the proper reference frame in $\mathbb{C}$, representing the $x$ and $y$ coordinates in $\mathbb{R}$ and the $z$ coordinate in $\mathbb{C}$. The system represented on the $x$-axis depends on the variable $\theta$ and the system represented on the $y$-axis depends on the variable $\alpha$.}
\label{Fig3}
\end{figure}

\begin{figure}[htb!]
\centering
\includegraphics[width=80mm]{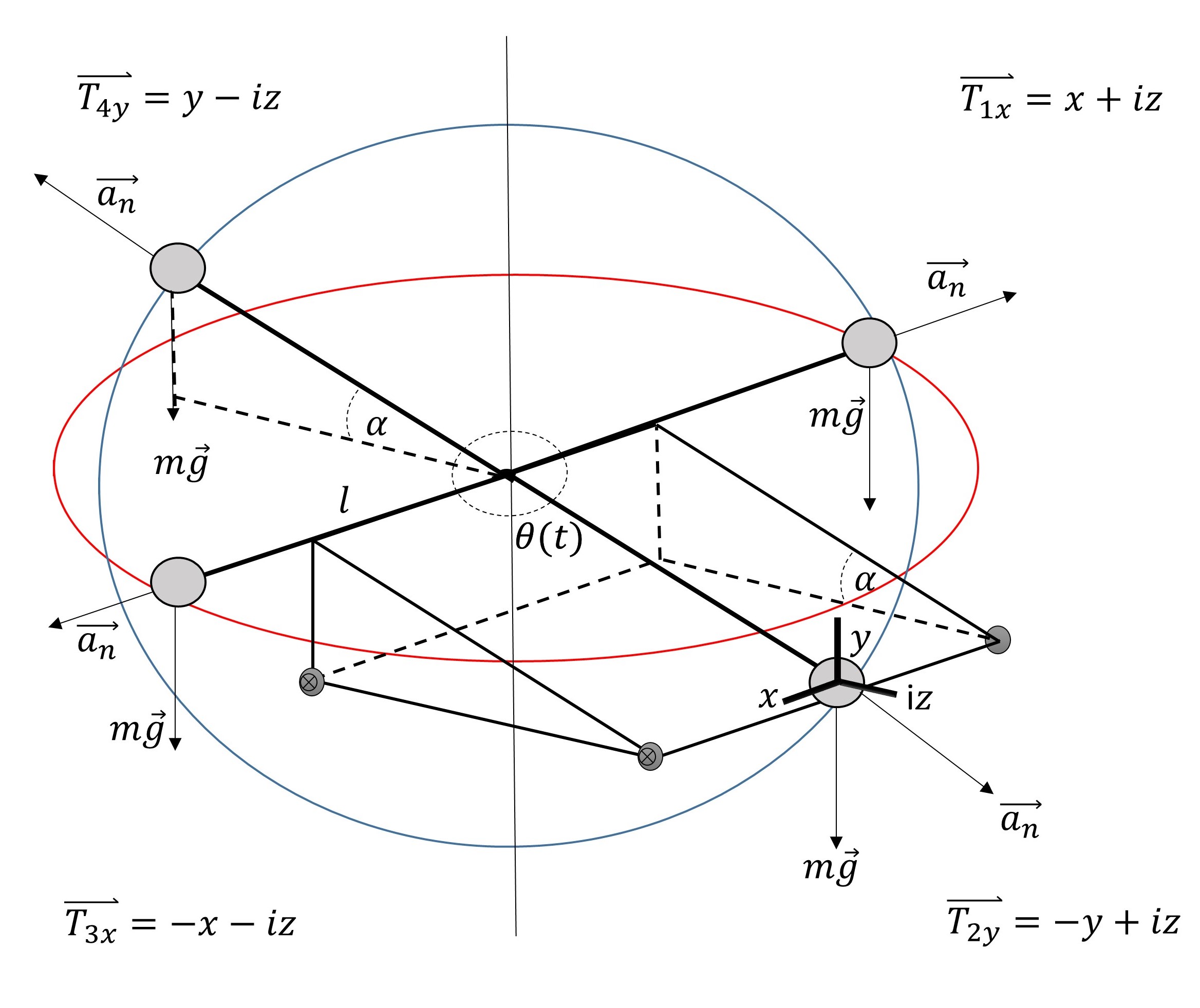}
\caption{Description of the proper reference frame in $\mathbb{C}$, representing both the $x$ and $y$ coordinates in $\mathbb{R}$ and the $z$ coordinate in $\mathbb{C}$. The system represented on the $x$-axis depends on the variable $\theta$ and the system represented on the $y$-axis depends on the variable $\alpha$.}
\label{Fig4}
\end{figure}

In order to represent in $\mathbb{C}$ a three-dimensional system it is necessary to superimpose two bidimensional planes that have in common the imaginary axis $z$ (Figs. \ref{Fig2}, \ref{Fig3} and \ref{Fig4}). This is a procedure for which there are no previous examples in the literature that has been consulted. In previous studies (paper 1), the forces governing the pendulum motion in the three Cartesian axes were described (Eq.\ref{eq:3}).

\begin{equation}\left.\begin{matrix}

\vec{F_{x}}=(m\vec{g} \sin\theta \sin\alpha)\;\widehat{x} \\ 
\hspace{15mm}
\vec{F_{y}}=(-ml\vec{\omega^{2}}\sin\alpha \cos\theta -m\vec{g})\;\widehat{y} \\
\hspace{15mm}
\vec{F_{z}}=((ml\vec{\omega^{2}} +m\vec{g}\cos\theta) \cos\alpha)\;\widehat{z}

\end{matrix}\right\}
\label{eq:3}
\end{equation} 
 
\hspace{5mm} 
 
It is possible to represent the components of the force of the pendulum of mass $m$, which are given by the weight and normal acceleration variables (Eq.\ref{eq:4}).

 \begin{equation}\left.\begin{matrix}
\vec{F_p}=mg\sin \theta \sin \alpha\;\widehat{x} -mg\;\widehat{y} +img\cos\theta\cos\alpha\;\widehat{z} \\

\vec{a}_n=-ml\vec{\omega^{2}}\sin\alpha\cos\theta\;\widehat{y}+iml\vec{\omega^{2}}\cos\alpha\;\widehat{z}  

\end{matrix}\right\}
\label{eq:4}
\end{equation}

%-------------------------------------------------------------------------
%                   Producto vectorial

%-----------------------------------------------------------------------

\subsection{Cross product}

Once the transformations of the system in complex numbers are known, it is possible to interpret the meaning of the displacement of the mobile due to the inertial force created by the pendular movement, With this aim, the scalar product will be calculated.

The cross product is defined as a binary operation in a three-dimensional space that produces another vector that is orthogonal to the two vectors that are multiplied \cite{hernandezproducto,mosquera2004magnitudes}.

%descripción del producto vectorial y qué tiene que ver con el desplazamiento del movil.

With the equations of weight and normal acceleration it is possible to know any value for the different angles $\alpha$ and $\theta$. The reference frame is centered at the bottom of the pendulum's motion, where the maximum of kinetic energy and the minimum of potential energy occur. In order to know the force vector (Eq.\ref{eq:5}) of the pendulum at each moment the cross vector provided by these two quantities will be represented.

\begin{equation}
    \vec{T_v}= \vec{F_p} \times  \vec{a}_n
    \label{eq:5}
\end{equation}

The initial point of the pendulum's motion is taken as the value where the potential energy will be maximum and the kinetic energy minimum, so the pendulum travels an angle of $180^{\circ}$ until it reaches the reference position described above. 

As explained, the motion of the system is possible only in the $z$-axis, given the arrangement of the wheels. Once the pendulum travels the first $90^{\circ}$, it begins to accumulate the force due to inertia, which is released, like a flywheel, when it is at the point of maximum kinetic energy or minimum potential energy ($180^{\circ}$). At that moment the mobile experiences a displacement in the positive $z$-axis.

In Equation \ref{eq:6} it can be observed that, when substituting the values of $m\vec{g}$ and $\vec{a_n}$ of Equation \ref{eq:4} in Equation \ref{eq:5} and making the cross vector, the coordinates that were in $\mathbb{R}$ become in $\mathbb{C}$ and viceversa.

\begin{equation}
 \vec{T_v}= m{g} \; {a}_n(i\cos\alpha(\sin\alpha   \cos^2\theta-1), -i\sin\alpha\cos\alpha \sin\theta, -\sin^2\alpha \sin\theta\cos\theta)
 \label{eq:6}
 \end{equation}

The effect of that permutation is that, although in the first instance there is no real magnitude in the direction of the $z$-axis, once the pendulum motion takes place the $z$-axis is the only one in which a real magnitude appears. The forces operating in the system can only give rise to a displacement when their component is projected in that direction. This is the reason why, if the system has wheels that allow it to move along the $z$-axis, the forces that govern the oscillatory motion of the pendulum can be transformed into a linear force of displacement of the mobile for the position of maximum kinetic energy.

%begin{equation}
%    \vec{T_v}= m\vec{g} \times  \vec{a_n}= mg %(sin(\theta)sin(\alpha), -1, icos(\theta)cos(\alpha)) %\times {a_n} (0, sin(\alpha)cos(\theta), icos(\alpha))
%\label{eq:5.2}
%\end{equation}

%-------------------------------------------------------------------------
%                  Producto escalar

%------------------------------------------------------------------------

\subsection{Dot product}

In contrast to the cross product, which results in a vector, the dot product of two vectors is called the scalar obtained as the product of the modulus of the vectors by the cosine of the angle they form.
Thus, for each value of $\alpha$ and $\theta$ the dot product of the force vectors corresponds to a point in the system, just as in the matrix (paper 2) each angular position defined by $\alpha$ and $\theta$ corresponds to an apparent weight. Each dot product of the system is corresponding to a single value (Eq.\ref{eq:7}).

\begin{equation}
    {T_e}= m\vec{g} \cdot  \vec{a}_n
    \label{eq:7}
\end{equation}

Once we know the equation that defines the dot product of the system (Eq.\ref{eq:8}), which results from substituting Eq.\ref{eq:4} in Eq.\ref{eq:7} and obtaining the dot product, it is possible to represent it for all the angles $\theta$ and $\alpha$, keeping in mind that for each value of the slope $\alpha$ 360 values of $\theta$ will be obtained. 

\begin{comment}
\begin{equation*}
  {T_e}=mga_n(\sin\alpha\cos\theta+i\cos\theta\cos\alpha i\cos\alpha)
\end{equation*}
\end{comment}

\begin{equation}
  {T_e}= mg\;{a_n}( \sin\alpha\cos\theta-\cos\theta\cos^2\alpha)
  \label{eq:8}
\end{equation}

\begin{figure}[htb!]
\centering
\includegraphics[width=0.8\linewidth]{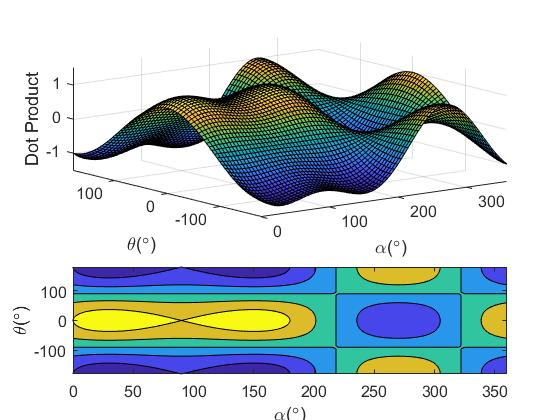}
\caption{Representation of the dot product for all the possible values of the angles $\theta$ and $\alpha$ within the system.}
\label{Fig5}
\end{figure}

It should also be noted, with regard to the unit in which the dot product is expressed, that since the starting units are $[\unit[]{m/s^2}]$ acceleration and $[\unit[]{kg~m/s^2}]$ force, the result of the dot product is given in $[\unit[]{kg~m/s^2}]$ or, in other words, $[\unit[]{W/s^4}]$, the result of the dot product is given in $[\unit[]{kg~m^2/s^4}]$ or, what is the same, $[\unit[]{W/s}]$, which could be expressed as a derivative of the power.
That is, for each dot product, calculated for each angle, variation of the power as a function of time is obtained. 

Although it might be thought that to know the behavior of the system for all values of $\theta$ and $\alpha$ it would be enough to calculate the first $180^{\circ}$ of each angle $\alpha$, the representation in Figure \ref{Fig5} shows that the system is not symmetric every $180^{\circ}$ but has a continuous symmetry every $360^{\circ}$.

\begin{figure}[htb!]
 \centering
  \subfloat[Plane representation seen from $\alpha$.]{
   \label{fig5.1}
    \includegraphics[width=0.4\textwidth]{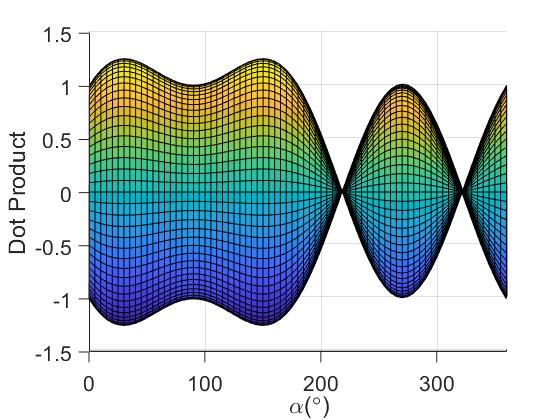}}
  \subfloat[Plane representation seen from $\theta$.]{
   \label{fig5.2}
    \includegraphics[width=0.4\textwidth]{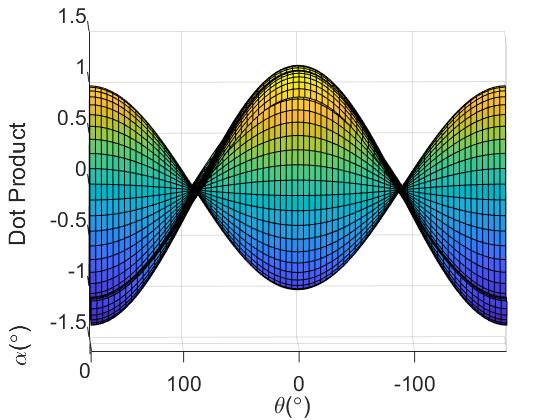}}
\\
 \subfloat[Representation of $\alpha$ for $\theta$ = $180^{\circ}$.]{
   \label{fig5.3}
    \includegraphics[width=0.4\textwidth]{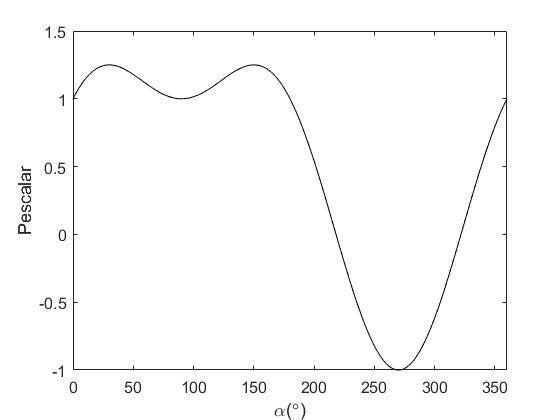}}
    \subfloat[Representation of $\theta$ for $\alpha$ = 36.37$^{\circ}$.]{
   \label{fig5.4}
    \includegraphics[width=0.4\textwidth]{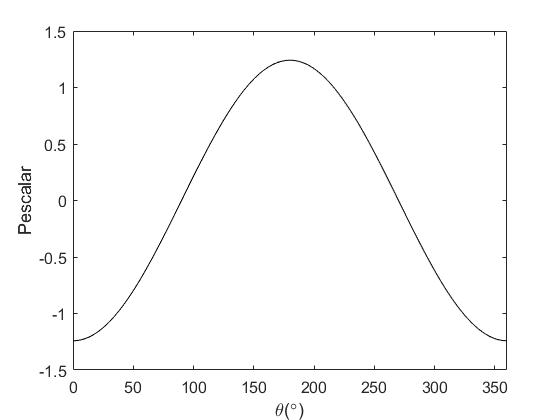}}
    
\caption{Dot product for the different angles $\theta$ and $\alpha$. a) View from the $\alpha$ plane. b) View from the $\theta$ plane. c) Representation of the cut in the $\alpha$ plane for $\alpha$ = $180^{\circ}$. d) Representation of the cut in the $\theta$ plane for $\alpha$ = 36.37$^{\circ}$. }
\label{fig5.x}
\end{figure}

By studying the angles for which the kinetic energy is maximum and the potential energy minimum, the representations in Figure \ref{fig5.x} are obtained. In Figures \ref{fig5.1} and \ref{fig5.3}, the profile that the dot product has as a function of $\alpha$ and viewed from $\theta$=$180^{\circ}$ is observed. The contour of the scalar product is represented in Figure \ref{fig5.3}, being able to visualize and verify two maxima, located at $30^{\circ}$ and $150^{\circ}$, and a minimum at $270^{\circ}$. 

On the other hand, Figures \ref{fig5.2} and \ref{fig5.4} show, respectively, the profile and contour obtained for the dot product as a function of $\theta$ and seen for $\alpha$=36.37$^{\circ}$ which shows a sinusoidal behavior.

Since in other works \cite{2021twodimensional} it was proved that the displacement of the mobile was maximum when the slope was 36.37$^{\circ}$, the transposition of the system to the imaginary plane makes it possible to show that in the period of the system, which covers $360^{\circ}$, there occur, at intervals of $120^{\circ}$, two maxima and a minimum, corresponding respectively to the angular values of $30^{\circ}$, $150^{\circ}$ and $270^{\circ}$. Although in the described mechanical system configuration the second maximum and minimum have no expression, given the resistance of the ground and the force of gravity, in other contextual assumptions (e.g. in vacuum or water) they would. 

\begin{figure}[htb!]
\centering
\includegraphics[width=50mm]{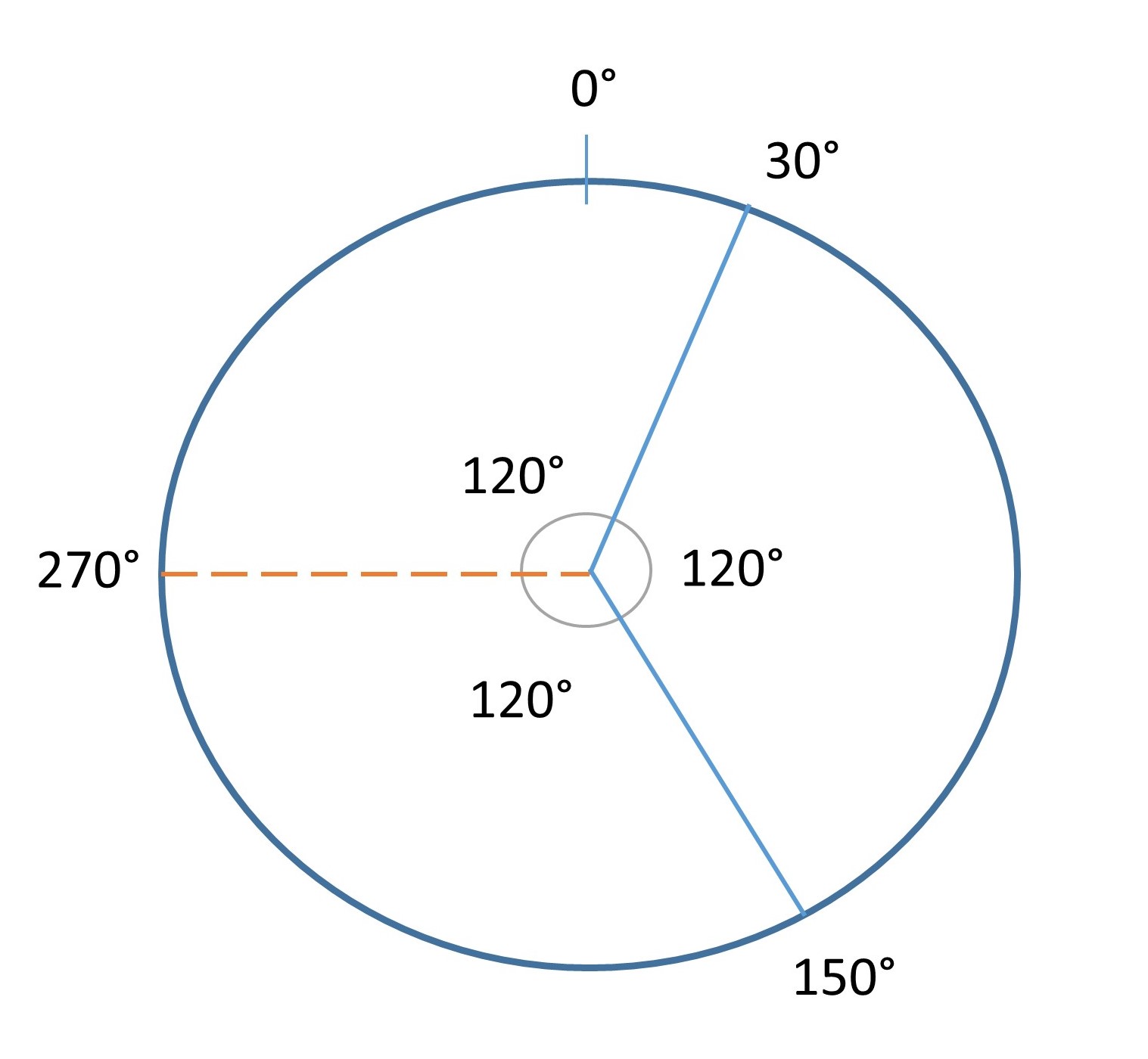}
\caption{Representation of the maximum kinetic energy for $\theta$ = $180^{\circ}$. The values of $\alpha$ maxima at $30^{\circ}$, $150^{\circ}$ and a minimum at $270^{\circ}$ are observed.}
\label{Fig6}
\end{figure}

The maximum and minimum angles for $\theta$ = $180^{\circ}$ observed in Figure \ref{fig5.3}, $\alpha$= $30^{\circ}$, $150^{\circ}$ and $270^{\circ}$, are plotted in Figure \ref{Fig6} to make visually evident that there is a difference of $120^{\circ}$ between them.

\section{Conclusions}

The use of a spherical coordinate system to determine the spatial position of the pendulum by means of a radius and two angles differs from the answer given in this article, since the space bounded in a spherical coordinate system does not include the space that contains the motion of the system that has been described. On the contrary, if the system is expressed in complex numbers, the set of real numbers is collected and all the roots of the polynomials are included.

Resorting to complex numbers to interpret the system allows to explain mathematically why the displacement of the mobile takes place even if there is no physical magnitude in the same direction and plane of the displacement. The performance of an operation such as the cross product between $m\vec{g}$ and $\vec{a}_n$ gives the representative character to the mobile system, since the physical magnitudes that begin in the complex plane are transformed into a real magnitude.

On the other hand, if the dot product of these same magnitudes is obtained on the system, a dot representation with units in $[\unit[]{W/s}]$ are obtained, and that the pendulum representation in a given time follows a sinusoidal path, as initially described, the system studied contains a certain slope, expressed in the complex plane, so that its scalar product shows a series of maxima and minima that could not be represented in the real plane. With this, a periodicity is observed in the system of $360^{\circ}$ representing the system in complex numbers. The absolute maxima and minima of the scalar product when the kinetic energy of the system is at an angular distance of $120^{\circ}$ from each other. As the minimum dot product is negative, one could decompose either of the two powers twice. This decomposition would be impossible mechanically, but it would be obtainable through electro-magnetical means or if the system was in a different medium.

\break

%\section*{References}

\bibliography{mybibfile}

\end{document}